# Earth's rotation forms the general circulation of the atmosphere.

Kochin A.V.

Central Aerological Observatory

3 Pervomayskaya str., Dolgoprudny, Moscow region, 141707 Russia

Email: amarl@mail.ru

**Abstract**

The general circulation of the atmosphere determines the long-term variability of weather processes. This circulation is driven by the temperature differences between the poles and the equator, causing air to move along the Earth's surface. However, this requires enhanced pressure at the poles, which is not observed. To sustain the circulation, an additional non-hydrostatic pressure gradient is required. In my research, I propose the emergence of an additional non-hydrostatic pressure gradient resulting from the centrifugal force generated by the Earth's rotation. This centrifugal force creates a non-hydrostatic vertical pressure gradient, which is essential for the closed circulation of unequally heated air in the meridional direction. The circulation is composed of three distinct streams flowing in opposite directions, with the polar and tropical tropopause acting as boundaries. The temperature in the atmosphere decreases from the surface to the polar tropopause and remains constant above it.

**Plain language summary**

Forecasting the characteristics of the general atmospheric circulation is important for long-term weather forecasting in order to ensure energy security. In 1753, the English scientist Hadley proposed a single-cell structure of the general atmospheric circulation, known as the convective cell, in which air rises at the equator and descends at the poles. However, temperature decrease should lead to increased surface pressure at the poles to drive the air masses along the Earth's surface from poles to equator. The observed surface pressure remains relatively constant, which means that the air can only move towards the poles due to the hydrostatic pressure gradient. To sustain the circulation, an additional non-hydrostatic pressure gradient is required. I consider the process of generating this additional non-hydrostatic pressure gradient through the centrifugal force resulting from the Earth's rotation. The centrifugal force balances the surface pressure and creates a non-hydrostatic vertical pressure gradient necessary for the closed circulation of unequally heated air in the meridional direction. The circulation consists of three oppositely directed streams separated by the polar and tropical tropopause, with the temperature in the atmosphere falling from the surface to the polar tropopause and constant above it.

## 1. Introduction

The study of atmospheric circulation has a rich historical background. As early as 1686, Edmond Halley published an article discussing the influence of solar heating on atmospheric motions. In 1753, Halley's work was further developed by Headley, who proposed a single-cell structure, known as a convective cell, for the general circulation of the atmosphere, with air rising at the equator and sinking at the poles. Numerous textbooks and monographs, such as Chamberlain (1978), Eckart (1960), Holton (2004), and Perevedentsev (2013), provide descriptions of the general circulation of the atmosphere. Despite the extensive research conducted over the years, our understanding of the physics underlying the processes in the general circulation of the atmosphere is still considered incomplete.

The general circulation of the atmosphere (GCA) plays a crucial role in transporting heat and moisture between the equator and the poles, thereby influencing long-term atmospheric processes. It is important to note that the speed of air flows in the GCA is significantly slower compared to the flows involved in mesoscale processes. Cyclones and anticyclones are responsible for more intense heat and water vapor transfer over the Earth's surface, but their effects are temporary and occur in multiple directions. In contrast, the GCA facilitates a permanent and unidirectional transfer, ensuring a sustained and consistent redistribution of heat and moisture across the globe.

The determination of the meridional velocity value is crucial for accurately estimating the circulation period of air masses in the GCA. However, the precise determination of this value is still an ongoing challenge. At the Earth's surface, where constant surface pressure is observed, the flow velocity within the GCA is practically negligible. To measure air flow speeds, observations are typically taken at a height of approximately 10 meters above the surface. Therefore, estimating the flow speed within the GCA based on ground-based meteorological observations is impossible. Additionally, the accuracy of upper-air data in measuring wind direction decreases when the flow velocity is less than 2 m/s. Furthermore, the presence of constant mesoscale processes in the atmosphere masks the flows within the GCA, given that the meridional component of the GCA is significantly smaller than the air velocity associated with these processes.

## 2. Formation of the general circulation of the atmosphere due to the vertical pressure gradient.

The general circulation of the atmosphere is primarily driven by the temperature contrast between the equator and the poles. In colder air, there is a greater vertical pressure gradient, resulting in a pressure difference between warm and cold regions that increases with altitude (Chamberlain 1978, Eckart 1960, Holton J. 2004, Perevedentsev 2013). In order for circulation to occur, it is essential that the atmosphere contains regions with varying directional gradients. As a result, the pressure near the Earth's surface in the vicinity of the poles should be greater than the pressure at the equator. This configuration prompts air near the surface to flow towards the equator, while above a certain height, the flow reverses with a zero gradient towards the poles. The surface pressure gradient required for the atmospheric circulation, driven by temperature variation, can be estimated assuming constant momentum flux. In the context of a stationary state within the general circulation of the atmosphere, it is assumed that the pressure integral should not depend on latitude to maintain the absence of a displacement of the atmospheric center of mass (Kochin 2020).

$\int_0^\infty P(h,\varphi)dh = const$ \qquad (1)

In an isothermal atmosphere, an interesting relationship emerges whereby the product of surface pressure $P_0(\varphi)$ and temperature $T(\varphi)$ remains constant.

$$P_0(\varphi)T(\varphi) = const \qquad (2)$$

In an isothermal atmosphere with a temperature difference of 60°K between the pole and the equator, and considering an equatorial temperature of 300°K and a surface pressure of 1000 HPa, the resulting pressure difference is 250 HPa. This pressure difference corresponds to a quarter of the total atmospheric pressure of 1000 HPa. In an isothermal atmosphere, the height of the constant pressure surface can be derived based on these conditions

$$h(\varphi) = \frac{R_c T(\varphi)}{g} \qquad (3)$$

h($\varphi$) is constant pressure surface height, $R_c$ is specific gas constant, g is free fall acceleration, T($\varphi$) is temperature at latitude $\varphi$. The height of the surface of constant pressure h($\varphi$) corresponds to the height scale in the atmosphere. Depending on the temperature, the height scale varies from 7 to 9 km. As an example, the dotted line in Fig. 1 represents the height of the tropopause during winter in the northern hemisphere (Gavrilova 1982), while the height scale H is combined with the thickness of planetary boundary layer (PBL ~ 1.5 km).

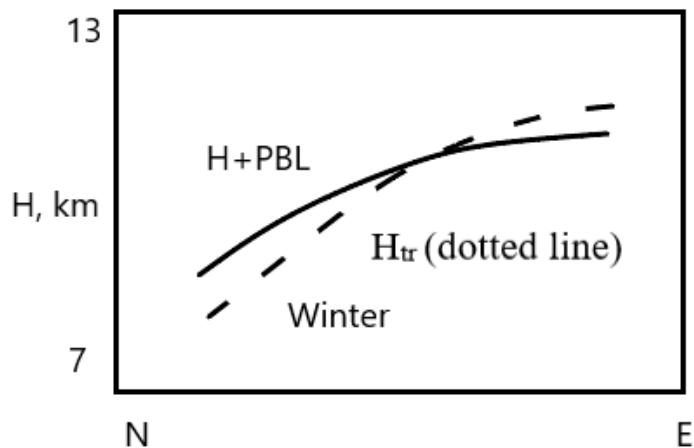

Fig. 1. Measured height of the polar tropopause during winter in the Northern Hemisphere (dotted line).

A strong correspondence between the height of the constant pressure surface and the tropopause height (Fig. 1) indicates that the tropopause serves as the boundary between meridional flows within the GCA. The model allows for the estimation of the meridional component velocity in the troposphere using the equation for viscous friction, which provides

$$U \sim \frac{dP}{2\mu L} \qquad (4)$$

where dP is the pressure difference, μ is the viscosity of air, and L is one-quarter of Earth's circumference. Assuming an air molecular viscosity of 15*10-6 Pa/s and a pressure difference of 250 HPa, the estimated velocity is approximately 100 m/s. However, turbulence plays a significant role in the troposphere, resulting in turbulent viscosity that is two to three orders of magnitude higher. Therefore, the estimated velocity of the meridional component in the troposphere ranges from 0.1 to 1 m/s. Considering the reduced mass of air above the tropopause, the estimated velocity in the stratosphere ranges from 0.3 to 3 m/s. The time required for air to travel from the pole to the equator is estimated to be at least 120 days, while from the equator to the pole, it is at least 40 days, resulting in a total transit time of approximately six months or more.

However, the model encounters a challenge in explaining the nearly constant surface pressure observed. In an isothermal atmosphere, the pressure-height relationship can be described by the equation

$$P(\varphi, h) = P_0 \exp\left(-\frac{gh}{R_c T(\varphi)}\right) \quad (5)$$

$P_0$ – surface pressure, h height above the surface, $R_c$ specific gas constant, $g$ free fall acceleration, $T(\varphi)$ temperature at latitude $\varphi$. In the case of a polytropic atmosphere, the relationship between surface pressure ($P_0$) and altitude (h) can be described by the equation:

$$P(\varphi, h) = P_0 \left(\frac{T_0(\varphi) - \gamma h}{T_0(\varphi)}\right)^{g/R_c \gamma} \quad (6)$$

$T_0(\varphi)$ surface temperature at latitude $\varphi$, $\gamma$ vertical temperature gradient. For an isothermal atmosphere, the derivative of the pressure $P(\varphi,h)$ in the atmosphere at height h and latitude $\varphi$ is $\frac{dP(\varphi,h)}{d\varphi} = \left(\frac{dP_0}{d\varphi} + \frac{dT(\varphi)}{d\varphi} \frac{P_0 gh}{RT^2(\varphi)}\right) \exp\left(-\frac{gh}{RT(\varphi)}\right) \quad (7)$

$P_0$ – surface pressure, $R_c$ specific gas constant, $g$ free fall acceleration, $T(\varphi)$ temperature at latitude $\varphi$. In an atmosphere with constant surface pressure, the first term in parentheses becomes zero, while the second term remains consistently negative across the entire atmospheric column due to the temperature T(φ) decreases from the equator to the pole. Air moves from an area of high pressure to an area of low pressure and due to hydrostatic pressure cannot move towards the equator. Thus, circulation cannot occur due to a hydrostatic pressure gradient and an additional non-hydrostatic pressure source is required for its occurrence.

### 3. Distribution of surface pressure due to the rotation of the Earth.

At the equator, the ocean level is approximately 21 km higher compared to the poles. This elevation difference is a result of the centripetal acceleration caused by the Earth's rotation. As a consequence, there is a pressure drop of approximately 2.1*10^6 Pa (pascals) in the ocean between the equator and the poles. This pressure drop can be likened to the rise of water along the walls of a rotating vessel. Similarly, centripetal acceleration influences on the atmosphere. If the atmosphere did not rotate with the Earth, the atmospheric pressure at the equator would be significantly lower, approximately 20 times less, than at the poles (as shown in Fig. 2).

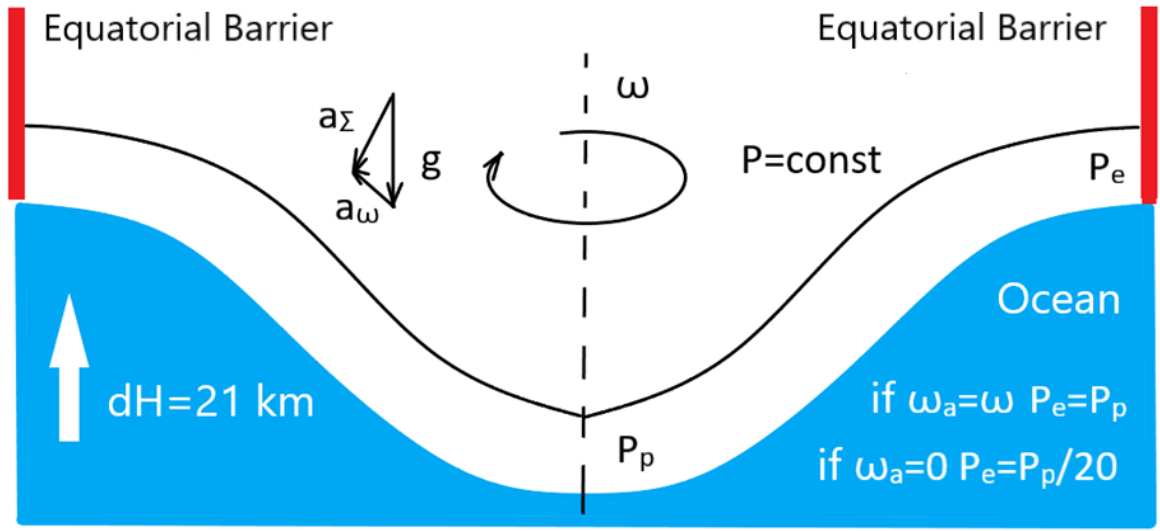

*Fig.2. Formation of surface pressure due to the rotation of the Earth.*

The centripetal acceleration, represented by $\omega^2 R$ contributes to an additional pressure gradient that drives air movement towards the equator, resulting in the formation of a constant pressure equipotential surface (Holton, 2004). This pressure gradient near the Earth's surface, which gives rise to a centripetal force, can be indirectly approximated by observing the pressure drop resulting from the difference in ocean surface heights between the poles and the equator.

$$\frac{dP}{dL} \sim \rho g \frac{2H}{\pi R} \approx 30 \; GPa \; /100 \; km \qquad (8)$$

Where ρ is the density of air near the Earth, 1.5 kg/m³, g is the acceleration of free fall, H is the difference in the heights of the ocean surface between the poles and the equator, R is the radius of the Earth. The pressure difference is three times the normal atmospheric pressure of 1000 HPa, and the pressure gradient is ten times larger than that observed in the largest extratropical cyclones with strong winds. As a result, the surface pressure remains constant due to the non-hydrostatic pressure gradient generated by the centripetal force. In the absence of temperature variations throughout the atmosphere, there is no hydrostatic pressure gradient in the meridional direction beyond the surface layer. This is because both terms in the equation (7) become zero. Consequently, the atmosphere reaches a state of equilibrium, and there is no meridional transport of air.

In a rotating vessel, the centripetal force causes an increase in pressure, pushing the liquid against the walls. Without walls, the liquid would flow out. In the atmosphere, a similar role is played by the Equatorial Barrier (shown in Fig. 2), which also exists in the ocean. At the equator, the centrifugal force changes its direction. The air mass from one hemisphere, due to the centripetal force, exerts pressure on the air mass from the other hemisphere. If a portion of air crosses the equator from any hemisphere, the centrifugal force from the other hemisphere will bring it back. As a result, the overall atmospheric pressure is equalized between the hemispheres. At the pole, the centrifugal force also changes direction, but its magnitude becomes zero due to a decrease in the perimeter.

Another important consideration is the impact of zonal velocity on surface pressure. The zonal component, denoted as V, is combined with the radial velocity of the Earth's surface, resulting in a modification of the angular rotation frequency. This change in the angular rotation frequency, referred to as $\omega_v$, can be calculated as follows

$$\omega_v = \frac{\omega_E R \pm V}{R} \quad (9)$$

Where $\omega_E$ is the angular frequency of the Earth's rotation, R is the radius at latitude φ, V is the zonal velocity. GCA describes meridional flows that move in opposite directions. When flowing from the pole to the equator, the zonal velocity is subtracted, whereas when flowing from the equator to the pole, it is added. It is likely that changes in zonal velocity within GCA do not directly impact surface pressure, but they can potentially influence meso- and macroscale processes, including the formation and behavior of powerful cyclones. It is important to acknowledge that the angular frequency of the Earth's rotation plays a significant role in shaping the distribution of air mass across its surface. Therefore, even slight variations in the angular frequency can have noticeable effects on the atmosphere (Sidorenkov, 2004).

## 4. Formation of circulation taking into account the centripetal force.

The occurrence of atmospheric circulation can be analyzed using a simplified model. In Figure 3, the processes are depicted both in the absence of the centripetal force (Fig. 3a and b) and in its presence (Fig. 3c and d).

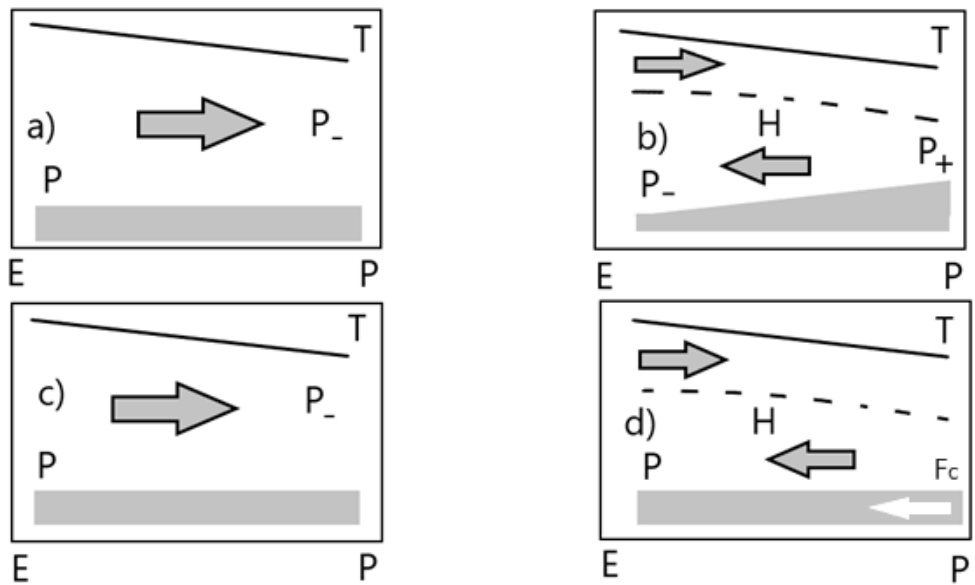

*Fig.3. Circulation formation.*

Prior to the formation of circulation, the atmospheric temperature is uniformly distributed, and the air mass is evenly spread across the surface. Surface pressure remains consistent everywhere. In the initial moment depicted in Figures 3a and 3c, the temperature of the atmosphere undergoes changes. Then the temperature at the pole decreases, which leads to the development of a temperature gradient between the pole and the equator, as shown by a solid line in Figure 3. To simplify matters, the temperature field is centered relative to the pole, with the minimum temperature coinciding with the pole. As the air cools, cold air compresses vertically, although the surface pressure remains uniform initially (indicated in gray). Near the surface, the air remains relatively stationary due to the small pressure difference caused by the vertical pressure gradient. However, above a certain height, the pressure gradient drives the air towards the pole, leading to an accumulation of air mass in that area. In the absence of a centripetal force, the pressure at the pole exceeds that at the equator, as depicted in gray in Figure 3b. Consequently, a surface pressure gradient forms, initiating a surface flow from the pole to the equator. At a certain height H (represented by the dashed line), a compensating flow is established due to the vertical temperature gradient, counterbalancing the surface flow.

In a rotating atmosphere, the presence of the centripetal force $F_c$ (indicated by the white arrow) prevents an increase in surface pressure, as it significantly outweighs the pressure difference attributed to temperature-related factors (equation 8). The temperature gradient induces poleward movement of the air, but near the surface, the centripetal force generates a reverse flow that works to equalize the pressure. Above a specific height H (represented by the dashed line), the poleward displacement of air gives rise to a compensating flow. It would seem that the occurrence of circulation has an explanation without contradiction with the difference in surface pressure. However, this model introduces a new challenge. The vertical distribution of pressure in such a scenario cannot align with the equation of hydrostatic equilibrium. According to the equation 8, the hydrostatic pressure at a constant surface pressure always decreases towards the pole.

To gain a better understanding of the circulation development, it is beneficial to examine the processes starting from the initial moment when the air mass is uniformly distributed, but the temperature has undergone changes. Temperature variations lead to differences in air density across the Earth's surface. At the equator, the air is less dense compared to the poles. This density disparity influences the vertical pressure gradient, which in turn generates a pressure difference causing the air to move. The air initiates movement towards the pole, and this displacement velocity increases with height, while remaining stagnant at the Earth's surface. Figure 4 illustrates the relationship between pressure difference and height, comparing the contributions from both the centripetal force (depicted by the red line) and the temperature gradient (represented by the blue line) in driving this circulation process.

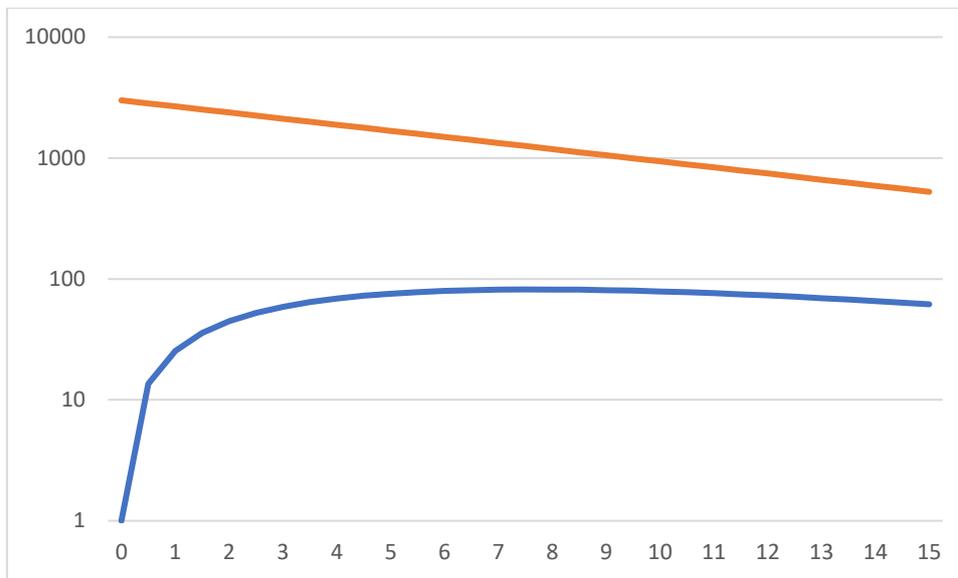

*Fig.4. Conditional dependence of the pressure difference on height between the equator and the pole due to the centripetal force (red line) and due to the temperature gradient (blue line). The horizontal axis represents the height in kilometers, while the vertical axis represents the pressure difference in HPa on a logarithmic scale. The pressure difference caused by the centripetal force is calculated based on the gradient magnitude derived from equation (8), accounting for variations in air density.*

The displacement of air to the poles does not directly occur at the surface but at a higher altitude. This is due to the dominance of centripetal force over the pressure gradient caused by temperature differences (Fig.4). While the displacement of air should theoretically result in an increase in surface pressure near the poles, the centripetal force keeps the surface pressure relatively constant. At the surface, the centripetal force creates a flow towards the equator, leading to a deficiency of air mass and a violation of hydrostatic equilibrium. This results in a smaller vertical pressure gradient than expected at a given temperature. Consequently, air descends towards the Earth's surface, forming a downward flow. This further enhances the surface flow towards the equator, increasing the vertical extent of the horizontal flow. The upstream flow from the equator to the pole reduces the upper atmospheric pressure near the equator, contributing to a decrease in surface pressure there. The centripetal force then acts to equalize the surface pressure near the equator by shifting an additional mass of air towards it. As a result, the surface pressure near the equator becomes higher than the hydrostatic pressure from the additional air mass present. This pressure difference leads to an upward vertical flow. The areas of these processes are approximately equal due to their similarity, so the boundary between the descending and ascending flows occurs at a latitude of approximately 30°. At this latitude, the descending flow replaces the ascending one, which looks like the boundary of the circulation cell. The descending and ascending streams are components of the general horizontal flow from the poles to the equator. Above them is a horizontal air flow from the equator to the pole. The velocities of the descending and ascending flow can be estimated from the law of conservation of mass. The horizontal mass transfer is equal to the flow between the boundaries of the regions with descending and ascending flows. The vertical flow flows in or out on the entire surface of the areas with descending and ascending flows. The vertical velocity can be estimated by the following expression

$$u_v \sim \frac{H}{R} u_h \quad (10)$$

Where $u_v$ is the velocity of the vertical flow, $u_h$ is the velocity of the horizontal flow, H is the height of the tropopause, R is the radius of the Earth. At a horizontal flow velocity of 1 m/s, the vertical flow velocity is about $10^{-3} – 10^{-4}$ m/s. It is challenging to detect the flow at such a speed. The structure of air flows is shown in Fig.5.

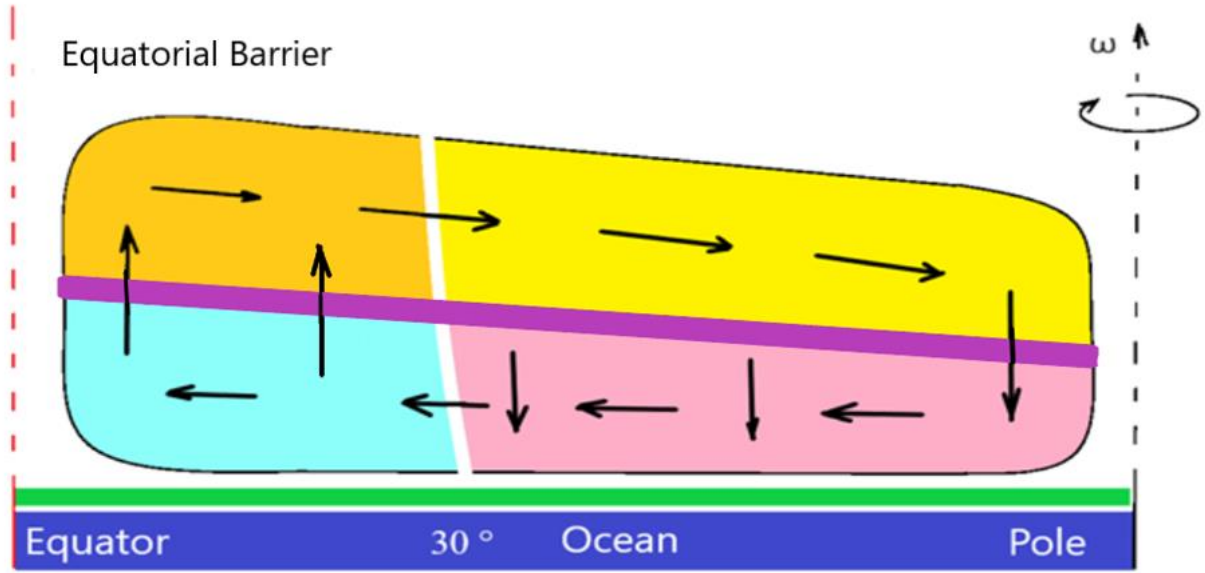

*Fig. 5. Air circulation in the GCA, taking into account the centripetal force. The green stripe is the planetary boundary layer, the purple line is the boundary between meridional flows. The downflow zone is highlighted in pink, the upflow zone is highlighted in blue, the compensating flow zone from the equator to the pole is highlighted in yellow.*

The flow structure depicted in Figure 5 is expected to exhibit specific temperature and wind speed profiles. In the descending flow, the temperature increases as the height above the surface decreases, primarily due to adiabatic compression during descent. As a result, for latitudes greater than 30°, the temperature should decrease with increasing altitude above the Earth's surface. Conversely, at latitudes less than 30°, the flow rises and cools, leading to a similar temperature profile. The meridional flows introduce a zonal component to the overall wind profile, influenced by the Coriolis force. Within the boundary layer of the atmosphere (highlighted in green in Figure 6), the impact of the pressure difference caused by the temperature gradient is minimal, and the GCA has a limited effect on processes, including temperature and wind profiles. The boundary between the meridional flows is characterized by an isothermal temperature profile since there are no ascending or descending flows in that region. The temperature gradient above this interface should be close to isothermal. The process of air descending along the interface (with decreasing height towards the pole at a ratio of 3) leads to heating and an overall increase in the temperature of the flow. Consequently, the upper part of the atmosphere above the interface becomes warmer towards the pole, but this does not alter the overall temperature gradient significantly. The ascent of air at latitudes less than 30°, driven by pressure gradients, can have an additional effect on the temperature gradient. However, due to simultaneous heating during descent along the interface, the temperature changes are relatively small. As a result, there should be a slight decrease in temperature with increasing altitude.

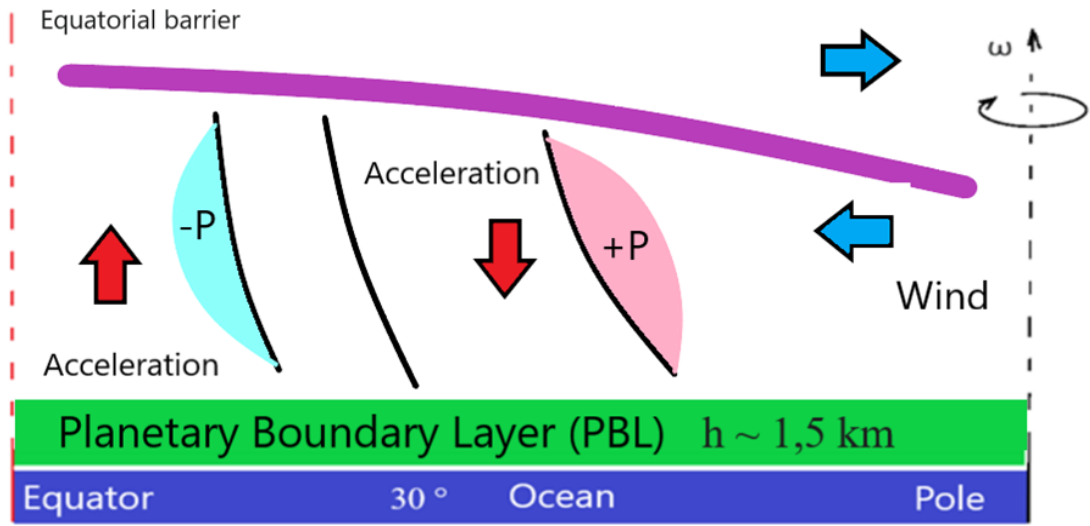

*Fig. 6. The occurrence of deviations from hydrostatic equilibrium. In the downflow zone, the air accelerates from the flow interface and decelerates at the upper boundary of the planetary boundary layer, which forms an area with increased pressure relative to hydrostatic equilibrium (highlighted in pink). In the zone of updrafts, air accelerates from the upper boundary of the planetary boundary layer and decelerates at the flow interface, which forms an area with reduced pressure relative to hydrostatic equilibrium (highlighted in blue).*

The occurrence of deviations from hydrostatic equilibrium is shown in Fig. 6. In the zone of descending flows, the air accelerates from the flow interface and slows down at the upper boundary of the planetary boundary layer, which forms an area with increased pressure relative to hydrostatic equilibrium (highlighted in pink). The upper boundary of the planetary boundary layer corresponds to a bend in the pressure difference due to the temperature gradient in Fig. 4 at an altitude of about units km. In the zone of ascending flows, the air accelerates from the upper boundary of the planetary boundary layer and slows down at the interface of the flows, which forms an area with reduced pressure relative to hydrostatic equilibrium (highlighted in blue).

Summing up the above, the temperature and wind profile should look like this (Fig. 7).

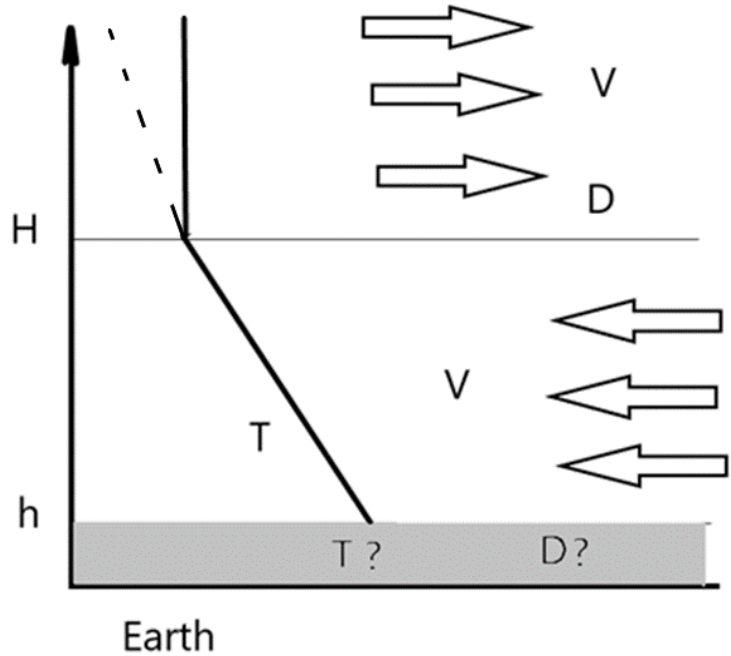

*Fig. 7. Vertical profile of temperature T and direction D of wind speed V. H is the separation height of the meridional flows, h is the height of the atmospheric boundary layer. The dotted line shows the temperature profile for latitudes less than 30°.*

The temperature profile in accordance with Fig.7 should be observed in the experimental data constantly. Detection of features in the zonal component of the wind is possible only under conditions close to calm in the atmosphere.

## 5. Experimental data on vertical temperature and wind profiles.

Detecting the meridional flow velocity in the GCA using current methods is challenging due to its relatively low magnitude, which is not prominently displayed in the wind field. Furthermore, this velocity is significantly smaller compared to speeds observed in mesoscale processes. Consequently, under conditions of relative atmospheric calm, only changes in wind direction within zonal flows can be detected. Figure 8 presents data from upper-air sounding, illustrating sharp wind direction changes occurring in both the polar and tropical tropopause regions.

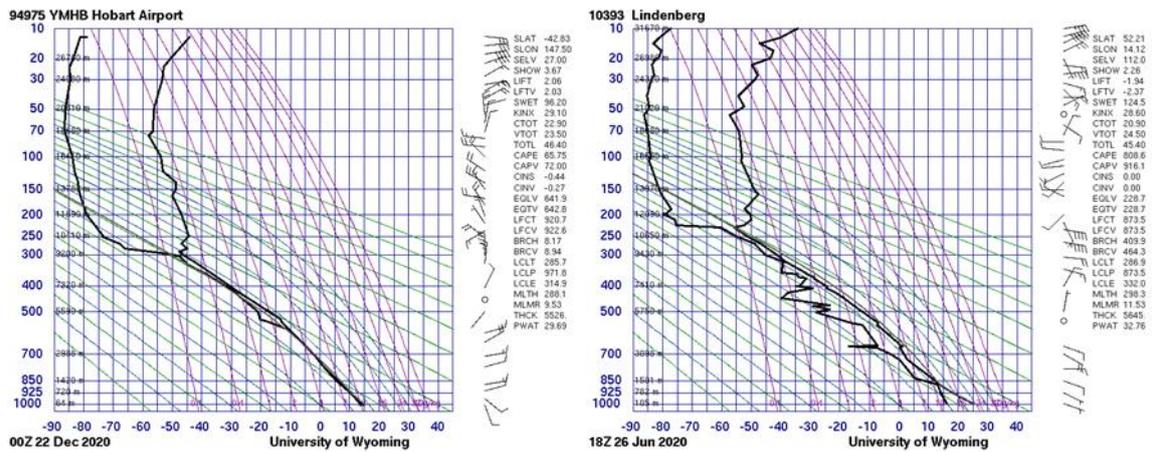

a) Hobart Airport, Australia, Lat: - 43º   b) Lindenberg, Germany, Lat: 52º

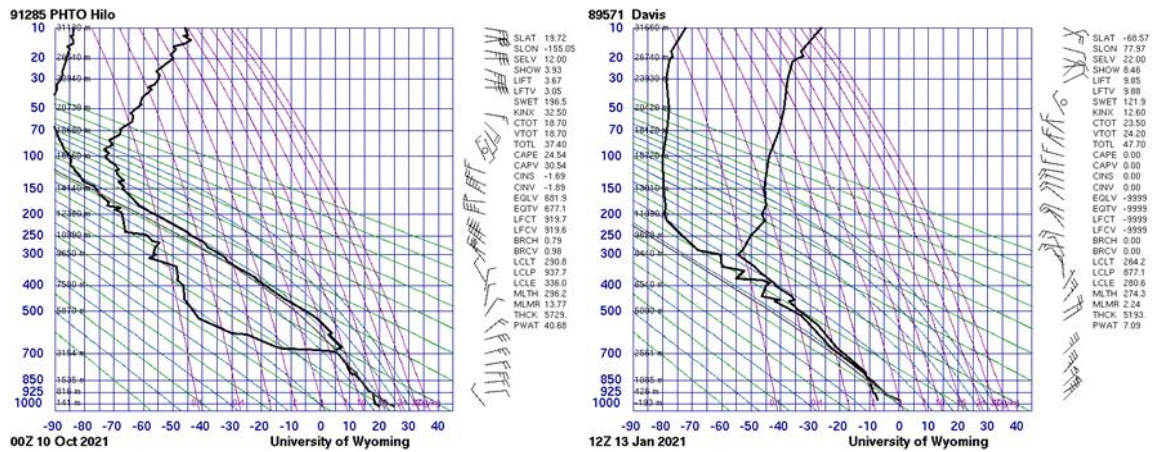

c) Hawaii, Latitude: 20º                               d) Davis, Antarctica, Lat: -69º

*Fig.8. Upper-air sounding results selected by the presence of a sharp change in wind direction near the tropopause. Wind direction is shown by arrows to the right of the charts. The arrow indicates where the wind is blowing. The air temperature is shown as a line closer to the center of the graph. The line on the left is the dew point temperature. The vertical axis shows altitude (m) and pressure (HPa), the horizontal axis shows temperature (Cº). The names of the stations and their latitudes are indicated below the charts (http://weather.uwyo.edu/upperair/sounding.html).*

From the Earth's surface to the tropopause, there is a predominant eastward transport of air (eastward flow). However, between the polar and tropical tropopauses, the flow direction reverses, resulting in a westward transport (western flow). Above the tropical tropopause, the air once again moves eastward (eastward transport). Consequently, the meridional transfer of air near the Earth's surface and above the tropical tropopause is directed towards the equator, while between the tropopauses, it is directed towards the pole. It turns out that there is a second circulation cell around the tropical tropopause, and the area between the tropopause is a hollow sphere that rotates from west to east. The temperature profile corresponds to the model profile in Fig. 8, including features at lo w latitudes (Fig. 8c).

Experimental data on the wind above the tropical tropopause are shown in Fig. 9 (with the permission of the author) and show similar changes at the stratopause level (Mariaccia 2023).

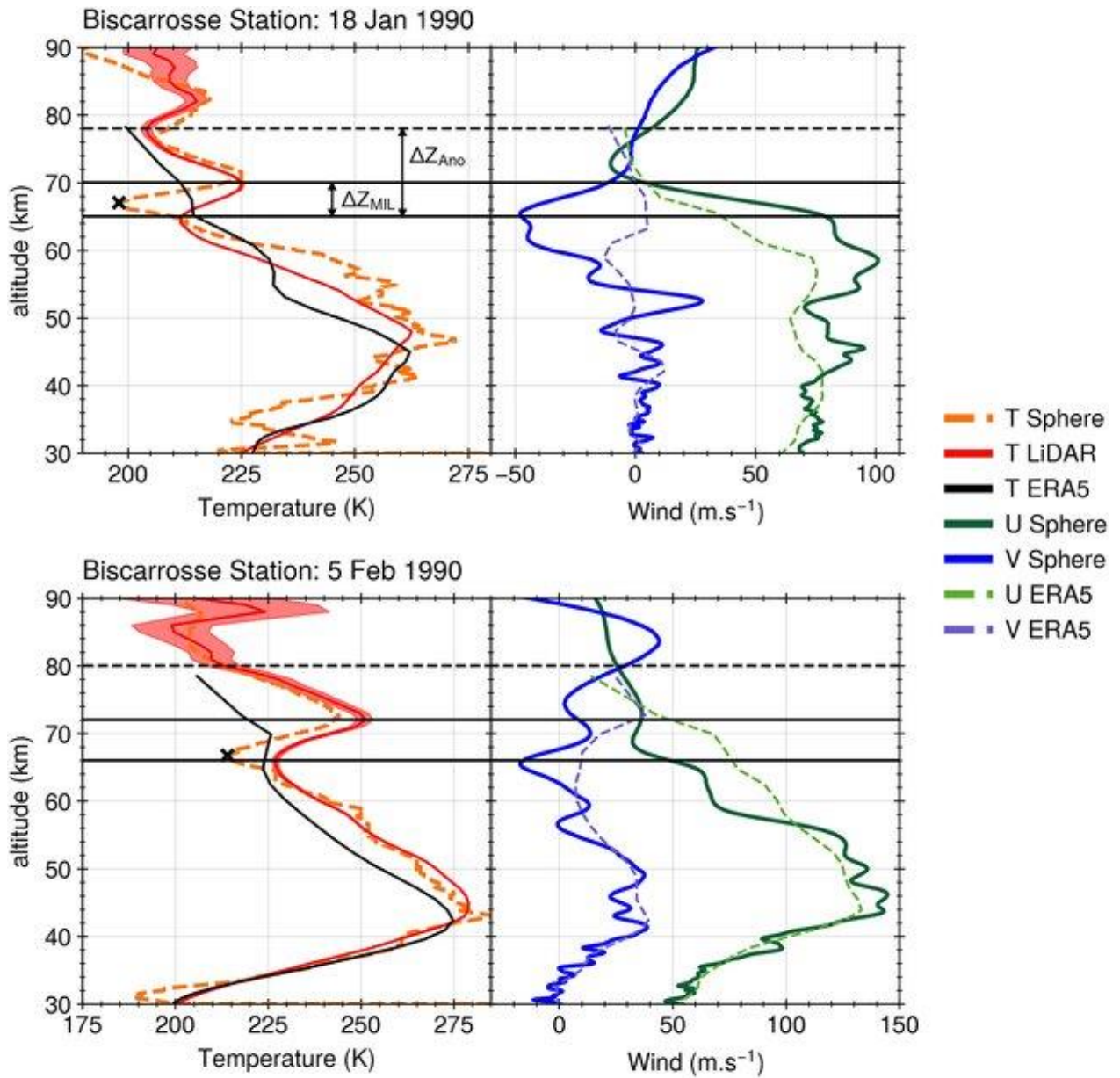

*Fig. 9. Temperature and wind profiles measured at Biscarrosse from falling spheres and Rayleigh LiDAR between 30 and 90 km for two dates during the DYANA Campaign. The statistical noise (red shaded area) of the LiDAR temperature signal is displayed. The two horizontal black solid lines indicate, respectively, the derived bottom and top of the MIL detected by the Rayleigh LiDAR. The horizontal dashed line represents the altitude of the potential total extension of the temperature anomaly (ΔZAno). The black cross points out the bottom of the MIL measured from falling spheres. In addition, the ERA5 temperature-wind profiles extracted for each date are shown.*

Hence, it is plausible to postulate the existence of an additional circulation cell in the atmosphere around the stratopause. This speculation is supported by the similarities in the wind behavior observed in the polar and tropical tropopauses, and the corresponding changes in the stratopause region.

# 6. Structure of flows in the general circulation of the atmosphere

The upper-air sounding data provides valuable insights into the overall atmospheric circulation (GCA) and suggests the presence of two distinct circulation cells. The formation of circulation around the polar tropopause can be attributed to the temperature gradient near the Earth's surface. The circulation around the tropical tropopause is influenced by the reversed temperature gradient in the stratosphere, as depicted in Figure 3 with a change in the temperature sign. As air descends along the interface, it undergoes adiabatic heating, resulting in a reversed temperature trend in comparison to the surface.

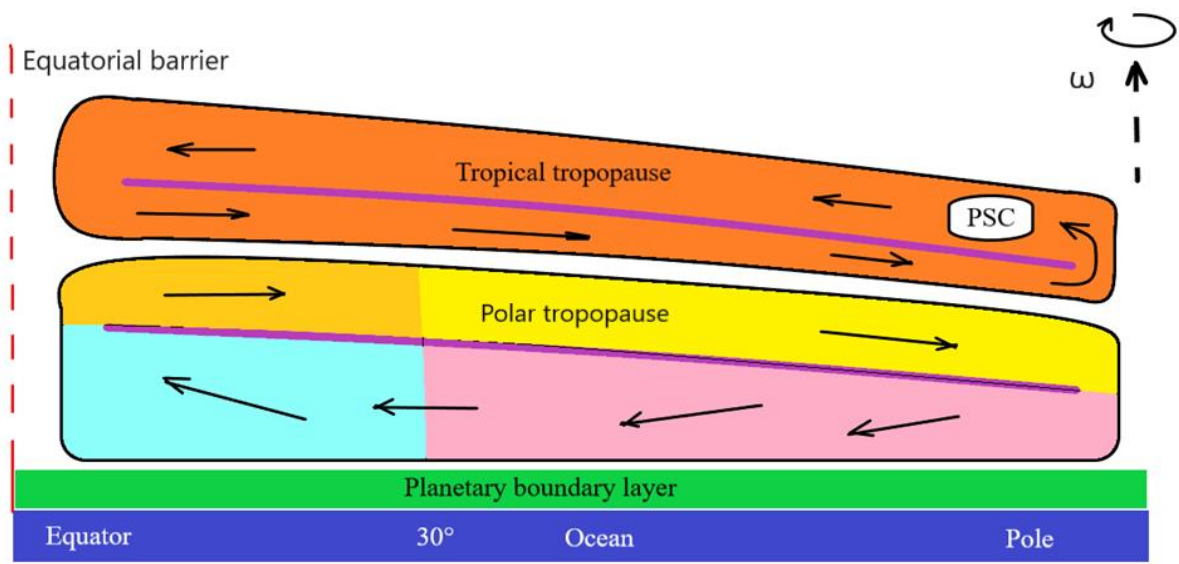

*Fig.10. The structure of flows in the general circulation of the atmosphere. The arrows show the movement of air. The tropopauses are shown as purple lines. The area with a downward flow and air movement from the pole to the equator is highlighted in pink. The region with an updraft and air movement from the pole to the equator is highlighted in blue. The area of flow from the equator to the pole is shown in yellow. The circulation around the tropical tropopause is shown in orange. Polar stratospheric clouds (PSC) are highlighted in white. The planetary boundary layer is shown in green.*

Figure 10 illustrates the flow structure within the global circulation of the atmosphere. The arrows indicate the direction of air movement, while the purple lines represent the location of the tropopauses. The region characterized by a downward flow and air movement from the pole to the equator is highlighted in pink, while the area with an updraft and air movement from the pole to the equator is highlighted in blue. The flow from the equator to the pole is depicted in yellow. The circulation around the tropical tropopause is shown in orange. Polar stratospheric clouds are highlighted in white, while the planetary boundary layer is depicted in green. The constancy of the tropopause heights is a result of the inertia of the zonal flows. Consequently, local temperature variations, including both daily fluctuations and seasonal changes, are effectively averaged out. Although large-scale temperature changes in the underlying surface can impact the nature of the general circulation, altering the speed of air flows, the fundamental structure remains unchanged. Within the tropopause, strong wind shears and jet streams can be observed due to the transition in wind direction.

Observational data confirms that there are distinct differences in air properties both up to the boundary of the circulatory flow section and above it. In the tropics, moist air rises and undergoes cooling, leading to the condensation of water vapor. Condensation ceases when the relative humidity reaches approximately 50-60%. Upon entering the stratosphere, the flow maintains a similar humidity level (e.g., as observed in Hawaii, as shown in Fig. 8c). However, the air temperature in the stratosphere is lower by 10-15 degrees compared to the middle latitudes (as depicted in Fig. 4a and Fig. 4b). While the actual water vapor pressure remains constant regardless of temperature, the saturated water vapor pressure increases significantly, which leads to a decrease in relative humidity. An increase in temperature by 10 degrees leads to a fourfold decrease in relative humidity, that is, it decreases to about 15%. With a temperature rise of 15 degrees, the relative humidity increases by seven times, resulting in a relative humidity level of only 8-9%. Consequently, a rapid change in relative humidity is observed at the tropopause.

In the lower atmosphere, up to the level of the polar tropopause, there is a significant concentration of aerosols. However, above the polar tropopause, the presence of aerosols is negligible, resulting in a sudden change in the extinction coefficient of visible light (Kochin 2021). Diffusion processes alone cannot account for the formation of such a distinct boundary with a vertical extent of approximately a hundred meters (as illustrated in Fig. 11).

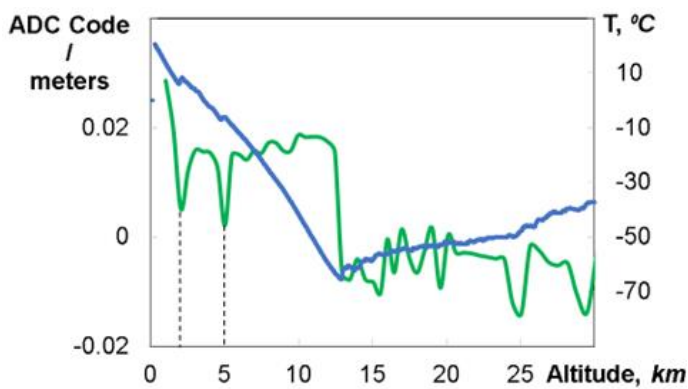

*Fig.11. Abrupt change in visible light extinction coefficient at the polar tropopause (Kochin 2021)*

The proposed structure of the GCA aligns with George Hadley's concept of a single-cell circulation between the equator and the poles. The boundary between the ascending and descending air currents is positioned at approximately 30° latitude, consistent with the traditional demarcation between the Hadley and Ferrell cells. In addition, the Ferrell cell at 30° latitude looks like a downdraft, which is also consistent with the proposed model. The presence of an independent circulation cell at temperate latitudes (Ferrell cell) implies a change in the direction of zonal flows, although this is not explicitly evident in the available observational data (as shown in Figs. 8a and 8c). It should be noted that wind data primarily indicate a shift in wind direction between the stratosphere and the troposphere, but provide limited information about the troposphere itself.

Detecting the presence of the Ferrell cell within the troposphere alone, if it is an independent cell only in this region, becomes challenging against the backdrop of wind changes attributed to mesoscale processes. The second boundary of the Ferrell cell is situated around 60° latitude, approximately coinciding with the latitude of the Arctic Circle (67°). However, the proposed model focuses solely on the scenario of a temperature minimum at the pole, which does not entirely correspond to the actual processes given the Earth's axial tilt of 23°. It is important to acknowledge that the existence of the Ferrell cell lacks consensus among researchers, with several scholars expressing doubts regarding its presence.

## 7. Conclusion and Implications

The GCA model presented in this paper so far represents only a qualitative description of the phenomena. However, its formulation based on the temperature gradient of the Earth's surface and the centripetal acceleration resulting from the Earth's rotation appears highly convincing due to its alignment with observational findings. Upper-air sounding data provides confirmation that the polar and tropical tropopauses act as boundaries between meridional flows moving in opposite directions. The consistent heights of the tropopauses can be attributed to the flow's inertia, which is influenced by the Coriolis force between these boundaries. A feature of the proposed GCA model is the formation of a vertical temperature profile with a decrease in temperature from the surface to the polar tropopause and a constant temperature above it.

Wind direction change (Mariaccia 2023) is also observed at the stratopause level. This may be due to the existence of another circulation cell in the atmosphere around the stratopause.

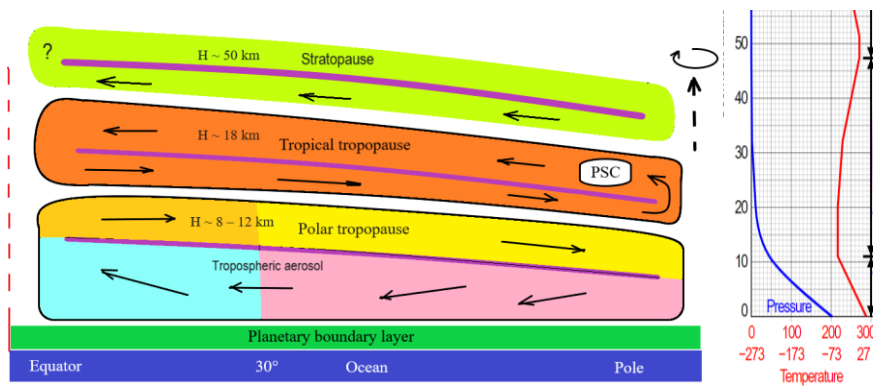

*Fig.12. Potentially possible general circulation in the atmosphere. The tropopause is shown by purple lines. PSC is polar stratospheric clouds. The temperature profile according to the International Standard Atmosphere is shown on the left.*

Potentially possible general circulation in the atmosphere is shown in Fig. 12. The tropopause is shown by purple lines. The direction of air circulation is shown by black lines. The polar tropopause blocks the flow of aerosol from the Earth's surface, which forms an aerosol layer in the troposphere. The rise of air at the pole in circulation around the tropical tropopause leads to the appearance of polar stratospheric clouds. The air circulation around the stratopause is still a hypothesis (shown in light green).

GCA plays a crucial role in the transfer of heat and moisture over the Earth's surface with a period of several months, which determines the long-term changes in weather phenomena. To develop a quantitative GSA model, it is necessary to obtain equations for describing the behavior of air in a rotating medium, taking into account the influence of the vertical temperature profile in the atmosphere and the influence of the Coriolis force on the flow velocity.

**Acknowledgments**

The author thanks the staff of the Central Aerological Observatory for their help and useful discussions during the work.